\documentclass[twocolumn,showpacs,prl,superscriptaddress]{revtex4}
\usepackage{graphicx}
\usepackage{psfrag}
\usepackage{amsmath} \usepackage{amsfonts} \usepackage{amssymb}

\begin{document}

\title{Spin waves in a one-dimensional spinor Bose gas}

\author{J.N. Fuchs}
\affiliation{Laboratoire de Physique des Solides, Universit\'e
  Paris-Sud, b\^at. 510, F-91405 Orsay, France }

\author{D.M. Gangardt}
\affiliation{\mbox{Laboratoire de Physique
Th\'eorique et Mod\`eles Statistiques,
Universit\'e Paris-Sud, b\^at. 100, F-91405 Orsay, France}}

\author{T. Keilmann}
\affiliation{Max-Planck-Institut f\"ur Quantenoptik,
Hans-Kopfermann Str. 1,
D-85748 Garching, Germany}

\author{G.V. Shlyapnikov}
\affiliation{\mbox{Laboratoire de Physique
Th\'eorique et Mod\`eles Statistiques,
Universit\'e Paris-Sud, b\^at. 100, F-91405 Orsay, France}}
\affiliation{\mbox{Van der Waals - Zeeman Institute, University of
  Amsterdam, Valckenierstraat
65/67, 1018 XE Amsterdam, The Netherlands}}

\date{\today}

\begin{abstract}
We study a one-dimensional (iso)spin $1/2$ Bose gas with repulsive
$\delta$-function interaction by the Bethe Ansatz method and discuss
the excitations above the polarized ground state. In addition to
phonons the system features spin waves with a quadratic dispersion.  We
compute analytically and numerically the effective mass of the spin
wave and show that the spin transport is greatly suppressed in the
strong coupling regime, where the isospin-density (or ``spin-charge'')
separation is maximal. Using a hydrodynamic approach, we study
spin excitations in a harmonically trapped system and discuss prospects
for future studies of two-component ultracold atomic gases.
\end{abstract}

\maketitle

Recent experiments have shown the possibility of studying ultra-cold
atomic gases confined in very elongated traps
\cite{1d_exp,PhilEss,Bloch,Weiss}. 
In such geometries, the gas behaves kinematically as if it were truly
one-dimensional (1D). Many theoretical studies
\cite{Girardeau,LL,Lieb,Olshanii,PSW,DLO} have predicted and discussed
interesting effects in 1D Bose gases, such as the occurrence of
fermionization in the strong coupling Tonks-Girardeau (TG) regime,
where elementary excitations are expected to be similar to those of a
non-interacting 1D Fermi gas \cite{Girardeau}. Manifestations of
strong interactions have been found in the experiments \cite{PhilEss},
and recently the TG regime has been achieved for bosons in an optical lattice \cite{Bloch}
and in the gas phase \cite{Weiss}.
 
Present facilities allow one to create spinor Bose gases which has
been demonstrated in experimental studies of two-component
Bose-Einstein condensates \cite{Cornell}. These systems are produced
by simultaneously trapping atoms in two internal states, which can be
referred to as (iso)spin 1/2 states. Relative spatial oscillations of
the two components can be viewed as spin waves
\cite{spin_dyn} (see \cite{Nikuni2003} for review).  A variety of
interesting spin-related effects such as phase separation
\cite{phasesep}, exotic ground states \cite{exotic}, and counter
intuitive spin dynamics \cite{spin_dyn} due to the exchange mean
field, have been studied both theoretically and experimentally.
However, most of these studies are restricted to the weakly
interacting Gross-Pitaevskii (GP) regime. There is a fundamental question
to what extent these effects survive in the strongly correlated regime
characteristic of one spatial dimension. The purpose of the present
Letter is to study spin excitations of an interacting 1D spinor Bose
gas. This is done using an exact solution by the Bethe Ansatz.

We start with a spinor (two component) gas of $N$ bosons with mass $m$
at zero temperature, interacting with each other via a repulsive
short-range potential in a narrow three-dimensional waveguide.  In
general, the interaction depends on the internal (spin) states
of the colliding particles. Here we consider the case of a
spin-independent interaction characterized by a single 3D scattering
length $a>0$.  This is a reasonable approximation for the commonly
used internal levels of $^{87}$Rb (see e.g. \cite{Cornell}).  The
waveguide has length $L$ and we assume periodic boundary conditions
for simplicity.  The transverse confinement is due to a harmonic
trapping potential of frequency $\omega_0$. When the chemical
potential of the gas is much smaller than $\hbar\omega_0$, the
transverse motion is frozen to zero point oscillations with amplitude
$l_0= \sqrt{\hbar/m\omega_0}$.  In such a quasi-1D geometry, the
interaction between atoms is characterized by an effective
one-dimensional delta-potential $g\delta(x)$. For $a\ll l_0$, the
coupling constant $g$ is related to the 3D scattering length as
$g=2\hbar^2 a/ml_0^2>0$ \cite{Olshanii}. The behavior of the system
depends crucially on the dimensionless parameter $\gamma = mg/\hbar^2
n$, where $n= N/L$ is the 1D density. For $\gamma\ll 1$ one has the
weak coupling GP regime, whereas for $\gamma\gg 1$ the gas enters the
strongly interacting TG regime.

Under the above conditions, the system is governed by the following
spin-independent 1D total Hamiltonian:
\begin{equation}
\label{eq:ham}
H=-\frac{\hbar^2}{2m}\sum_{i=1}^{N}\frac{\partial^2}{\partial
x_i^2}+g\sum_{i<j}\delta(x_i-x_j).
\end{equation}
This Hamiltonian was introduced by Lieb and Liniger \cite{LL} for
describing spinless bosons, and their solution by the Bethe Ansatz
(BA) has been generalized to bosons or fermions in two internal states
by M.~Gaudin and C.N.~Yang \cite{Gaudin,Yang}. In the case of a
two-component Bose gas (spin 1/2 bosons), due to the $SU(2)$
symmetry of the Hamiltonian the eigenstates are classified according
to their total (iso)spin $S$ ranging from 0 to $N/2$. 
In this case, which was recently considered by Li, Gu, Yang and 
Eckern \cite{LGYE}, the ground state is fully polarized ($S=N/2$) 
and has $2S+1$-fold degeneracy, in agreement
with a general theorem \cite{EL,YL}.  
At a fixed $S=N/2$, the system is described by the 
Lieb-Liniger (LL) model \cite{LL}, for which elementary excitations
have been  studied in Ref.~\cite{Lieb} for any value of the
interaction constant. Spin excitations above the ground state are
independent of the ground-state spin projection $M_S$ and represent
transverse spin waves. For $M_S=0$ they correspond to relative oscillations 
of the two gas components.  

We first give a brief summary of the BA diagonalization
\cite{Gaudin,Yang,LGYE} of the Hamiltonian (\ref{eq:ham}).  An
eigenstate with total spin $S=N/2-K$ ($0\leq K\leq N/2$) is characterized by two sets of
quantum numbers: $N$ density quantum numbers $I_j$ with $j=1,..,N$ and
$K$ spin quantum numbers $J_{\mu}$ with $\mu=1,..,K$.  If $N-K$ is odd
(resp. even), $I_j$ and $J_\mu$ are integers (resp. half-integers).
These quantum numbers define $N$ quasi-momenta $k_j$ and $K$ spin
rapidities $\lambda_{\mu}$, which satisfy the following set of BA
equations (we set $\hbar=2m=1$):
\begin{eqnarray}
\!\!\frac{L k_j}{2}\!\!&=&\!\!\pi I_j\!-\!\sum_{l=1}^{N}\!\arctan
     \!\!\left(\!\frac{k_j\!-\!k_l}{g/2}\!\!\right) \!+\!
     \sum_{\nu=1}^{K}\!\arctan\!\!\left(\!\frac{k_j\!-\!\lambda_\nu}{g/4}\!\!\right),
     \label{BAE1}\\
\!\!\pi J_\mu\!\! &=& \!\! \sum_{l=1}^{N}\!\arctan
     \!\left(\!\frac{\lambda_\mu\!-\!k_l}{g/4}\!\!\right)
     -\sum_{\nu=1}^K\!\arctan\!
     \left(\!\frac{\lambda_\mu\!-\!\lambda_\nu}{g/2}\!\!\right).
\label{BAE2}
\end{eqnarray}
The energy of the corresponding state is $E=\sum_{j} k_j^2$, and its
momentum is given by:
\begin{equation}
p=\sum_{j=1}^{N} k_j=\frac{2\pi}{L}\Big(\sum_{j=1}^N I_j
-\sum_{\mu=1}^K J_{\mu}\Big) .
\label{mom}
\end{equation}
As we are also interested in finite size effects, we do not take the
thermodynamic limit at this point.

The ground state corresponds to the quantum numbers
$\{I_j^{0}\}=\{-(N-1)/2,..,(N-1)/2\}$ and $K=0$, which shows that the
BA equations reduce to those of LL \cite{LL}. The wave function is
given by the orbital wave function of the LL ground state multiplied
by a fully polarized spin wave function. All ground state orbital
properties (energy, chemical potential, correlation functions, etc.)
are therefore identical to those of the LL model.  Elementary
excitations in the density sector correspond to modifying the density
quantum numbers $I_j$ while leaving the total spin unchanged, i.e.,
$K=0$.  At low energy, the density excitations are phonons propagating
with the Bogoliubov sound velocity $v_s=\sqrt{2gn}$ in the GP limit and with the
Fermi velocity $v_s=2\pi n$ in the TG regime.

We now focus on the spin sector. Elementary spin excitations
correspond to reversing one spin ($K=1$), and the total spin changes from
$N/2$ to $N/2-1$. Thus, we have a single spin rapidity $\lambda$ and the
corresponding quantum number $J$. In general, this
procedure creates a density excitation and a spin wave (isospinon)
\cite{LGYE}. Here, we choose specific quantum numbers $I_j, J$ in order to
excite the isospinon alone \cite{footnote}. Accordingly, the momentum $p$ of
the excitation is
\begin{equation}
p=\frac{2\pi}{L}\left(\frac{N}{2}-J\right),
\label{pJ}
\end{equation}
which follows from the definition (\ref{mom}).

In the long wavelength limit, where $|p|\ll n$, due to the $SU(2)$ symmetry
one expects \cite{Halperin} a quadratic dispersion for the spin-wave excitations
above the ferromagnetic ground state:
\begin{equation}
\varepsilon_p \equiv E(p)-E_0\simeq p^2/2m^{*}
\label{defiem},
\end{equation}
where $E(p)$ is the energy of the system in the presence of a spin wave with
momentum $p$, $E_0$ is the ground state energy and $m^{*}$ is an
effective mass (or inverse spin stiffness). This quadratic
behavior is due to a  vanishing  inverse spin susceptibility, which is
a consequence of the $SU(2)$ symmetry \cite{Halperin}.  A variational
calculation in the spirit of Feynman's single mode approximation
\cite{YL}, shows that $\varepsilon_p \leq p^2/2m$ implying that
$m^{*}\geq m$. Below we show that strong interactions greatly
enhance the effective mass.

In the strong coupling limit it is possible to solve the BA equations
(\ref{BAE1}) and (\ref{BAE2}) perturbatively in $1/\gamma$ \cite{LL}.  We
solve these equations both for the ground state $\{I_j^{0}\}$ and the
excited state $\{I_j;\, J\}$.  We anticipate that in the limit of strong
interactions, for small momenta ($|p|/n\ll 1$) and a large number of
particles ($N\gg 1$), the
dimensionless spin rapidity is $\tilde{\lambda}\equiv 2\lambda/ g \gg 1$
and the dimensionless quasi-momenta are $|k_j|/g\ll 1$. This allows us to
expand Eqs.~(\ref{BAE1}) and (\ref{BAE2}) to first order in
$1/\gamma$ and $1/N$. The ground state quasi-momenta are then given by:
\begin{eqnarray}
k_j^0 L=2\pi I_j^{0}\left(1-2/\gamma\right).
\label{k0}
\end{eqnarray}
Here we used the relation $\sum_l \arctan(2(k_j^{0}-k_l^{0})/g)\simeq
2Nk_j^0/g-2(\sum_l k_l^0)/g=2Nk_j^0/g$, which is a consequence of the vanishing
ground state momentum. Similarly, the excited state quasi-momenta
obey the equations:
\begin{eqnarray}
k_j L = \left(1-\frac{2}{\gamma}\right)2\pi I_j +\frac{2pL}{N\gamma}
-\pi+\frac{1}{\tilde{\lambda}}\left(1+\frac{k_jL}{\gamma N
\tilde{\lambda}} \right),
\label{k}
\end{eqnarray}
where $p$ is given by Eq.~(\ref{mom}). Neglecting quasi-momenta $k_l$ 
in the argument of arctangent in the BA equation (\ref{BAE2}), we obtain 
the excited state spin rapidity:
\begin{equation}
2\pi J=2N\arctan(2\tilde{\lambda})\simeq \pi N -N/\tilde{\lambda}
\label{J}
\end{equation}
Equations (\ref{pJ}) and (\ref{J}) then give:
\begin{equation}
\tilde{\lambda}=N/pL,
\label{lambda}
\end{equation}
which justifies that $\tilde{\lambda} \gg 1$ for $|p|/n \ll 1$. Combining
this result with Eq.~(\ref{k}) shows that $|k_j|/g \ll 1$, as
anticipated.  Let us now define the
shift of the quasi-momenta $\Delta k_j \equiv k_j-k_j^{0}$. 
Taking the difference between equations
(\ref{k}) and (\ref{k0}), we find:
\begin{equation}
\Delta k_j=\frac{1}{L \tilde{\lambda}}+\frac{k_j^{0}}{\gamma N
\tilde{\lambda}^2} +\frac{2p}{\gamma N}-\frac{2\pi}{L \gamma}
\label{Dkj}
\end{equation}
where we used that $I_j-I_j^{0}=1/2$. We can now compute the
energy of the spin wave, as defined in Eq.~(\ref{defiem}):
\begin{equation}
\varepsilon_p= \sum_{j=1}^{N} \left[ 2k_j^{0} \Delta k_j + (\Delta
k_j)^2 \right].
\end{equation}
Using Eq.~(\ref{Dkj}) for $\Delta k_j$ and Eq.~(\ref{lambda}) for 
$\tilde{\lambda}$ gives
$\varepsilon_p=p^2\left(1/N+2\pi^2/3\gamma \right)$.  Note that the
last two terms in the right hand side of Eq.~(\ref{Dkj}) give no
contribution, as the ground state momentum is zero. According to the
definition (\ref{defiem}), the inverse effective mass is therefore:
\begin{equation}
\frac{m}{m^{*}}=\frac{1}{N}+\frac{2\pi^2}{3\gamma},
\label{strongres}
\end{equation}
where we restored the units. Remarkably, the effective mass reaches the 
total mass $Nm$ for $\gamma\to \infty$: the bosons are impenetrable and
therefore a down spin boson can move on a ring only if all other bosons
move as well.

In the opposite limit of weak interactions it is possible to compute
the effective mass from the Bogoliubov approach \cite{LP}. The validity
of this procedure when considering a 1D Bose gas, i.e. in the absence
of a true Bose-Einstein condensate, is justified in \cite{Popov}.  The
Hamiltonian of the system can be written as $H_0+H_{int}$, where $H_0$ is
the Hamiltonian of free Bogoliubov quasiparticles and free spin waves:
\begin{equation}
H_0=\sum_{p}\epsilon_p \alpha_{p}^{\dagger}\alpha_{p}
+\sum_{p}e_p \beta_{p}^{\dagger}\beta_{p},
\end{equation}
with $\alpha_p,\beta_p$ being the Bogoliubov quasiparticle and the spin wave
field operators, $\epsilon_p = \sqrt{e_p(e_p+2gn)}$ the Bogoliubov spectrum, and
$e_p = p^2/2m$ the spectrum of free spin waves \cite{footnote2}. The Hamiltonian
$H_{int}$ describes the interaction between Bogoliubov quasiparticles and spin waves
and provides corrections to the dispersion relations $\epsilon_p$ and $e_p$. 
The most important part of $H_{int}$ reads:
\begin{equation}
H_{int}=g\sqrt{\frac{n}{L}}\sum_{k,q\neq 0}
\left(u_q\alpha_{q}^{\dagger}-v_q\alpha_{-q}\right)
\beta_{k-q}^{\dagger}\beta_{k}+\text{h.c.},
\end{equation}
where $u_q$ and $v_q$ are the $u,v$ Bogoliubov coefficients 
satisfying the relations $u_q+v_q=\sqrt{\epsilon_q/e_q}$ and
$u_q-v_q=\sqrt{e_q/\epsilon_q}$ \cite{LP}. Neglected terms
contribute only to higher orders in the
coupling constant. To second order in perturbation theory, in the 
thermodynamic limit the presence of a spin wave changes the energy of the system
by:
\begin{equation}   \label{prec}
\Delta E(p)=e_p +\frac{g^2n}{2\pi\hbar}\int dq\;
\frac{e_q}{\epsilon_q}\; \frac{1}{e_p-[\epsilon_q+e_{p+q}]}.
\end{equation}
In order to calculate a correction to the
effective mass of the spin wave, we expand Eq.~(\ref{prec}) in
the limit of $p\rightarrow 0$. Terms which do not depend
on $p$ modify the ground state energy, linear terms vanish, and
quadratic terms modify the spin wave spectrum as follows:
\begin{equation}
\varepsilon_p=e_p \left(1-\frac{4g^2 n}{\pi\hbar}\int_{0}^{\infty}
dq\; \frac{e_q}{\epsilon_q}\; \frac{e_q}{[\epsilon_q+e_q]^3} \right),
\end{equation}
where the main contribution to the integral comes from momenta $q\sim
\sqrt{mgn}$.  Using the
definition (\ref{defiem}), we then obtain the inverse effective mass:
\begin{equation}
  \frac{m}{m^{*}}=1-\frac{2\sqrt{\gamma}}{\pi}\int_0^\infty
  dx\,\frac{(\sqrt{1+x^2}-x)^3}{\sqrt{1+x^2}}
  =1-\frac{2\sqrt{\gamma}}{3\pi},
\label{weakres}
\end{equation}
which clearly shows non-analytical corrections to the bare mass due to
correlations between particles.  This result can also be obtained
directly from the BA equations.

\begin{figure}[]
\psfrag{m}{\large$\frac{m}{m^*}$} \psfrag{l}{\large$\log\gamma$}
\centering \includegraphics[width=8cm]{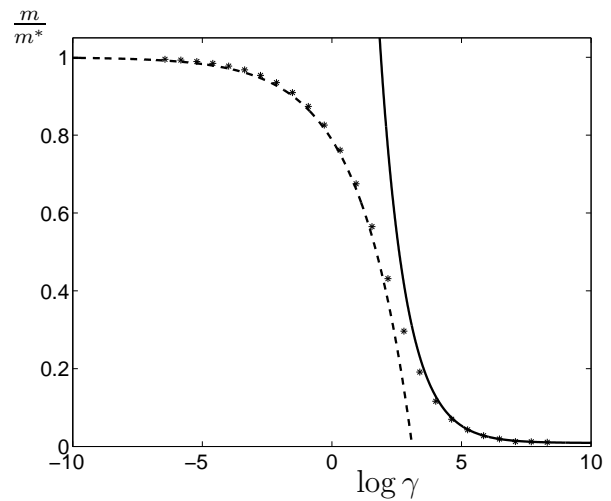}
\caption{Inverse effective mass $m/m^*$ as a function
of the dimensionless coupling constant $\gamma$ (logarithmic scale). The stars
($*$) show numerical results for $N=111$ particles, the solid curve
represents the behavior in the strong coupling limit (Eq.~(\ref{strongres})),
and the dashed curve the behavior for a weak coupling (Eq.~(\ref{weakres})). }
\label{effmass111}
\end{figure}
For intermediate couplings, we obtained the effective mass by numerically 
solving the BA equations (\ref{BAE1}) and (\ref{BAE2}). Our results 
are displayed in Fig.\ref{effmass111}. Note that when 
solving the BA equations, one should take care of
choosing $N^{-2}\ll \gamma \ll N^2$. Indeed, if $\gamma < N^{-2}$, the
potential energy per particle in the weak coupling limit is lower than
the zero point kinetic energy $\hbar^2/m L^2$ and the gas is therefore
non-interacting (effectively $\gamma=0$). In the strong coupling limit
and for the same reason, if $\gamma>N^2$, the system behaves as a TG
gas (effectively $\gamma=\infty$). 

We now turn to harmonically
trapped bosons in the TG regime and rely on spin hydrodynamics
introduced for uniform systems \cite{Halperin}. As the ground state is fully polarized we assume
the equilibrium (longitudinal) spin density $\vec{S} (x)=n(x)\hat{e}_3$ and
study small transverse spin density fluctuations $\delta\vec{
S}(x,t)=\delta S_1\hat{e}_1+\delta S_2\hat{e}_2 $, where $\hat{e}_1,\hat{e}_2,\hat{e}_3$ 
form an orthonormal basis in the spin space. For a large $N$,
the equilibrium density profile $n(x)$ in a harmonic trapping potential 
$V(x)=m\omega^2x^2/2$ is given by the Thomas-Fermi expression
 \begin{equation}
   \label{eq:dens_prof}
   n(x)=n_0\sqrt{1-(x/R)^2}.
 \end{equation}
Here $n_0=n(0)$ is the density in the center of the trap and
$R=\sqrt{2\hbar N/m\omega}$ is the Thomas-Fermi radius. For a strong
but finite coupling Eq.~(\ref{eq:dens_prof}) represents the leading term, with
corrections proportional to inverse powers of $\gamma_0=mg/\hbar^2
n_0$.  The spin density fluctuations $\delta\vec{ S}$ obey the
following linearized Landau-Lifshitz equations \cite{Halperin}:
\begin{equation}
  \delta \dot{S}_{1,2}=\mp\frac{\hbar}{2}\partial_x\frac{
    n(x)}{m^*(x)} \partial_x \frac{\delta S_{2,1}}{n(x)}
    .
\label{eq:hydro_spin}
\end{equation}
In the TG regime  the effective mass entering 
the equation of motion
(\ref{eq:hydro_spin}) depends on the density profile $n(x)$ as
 \begin{equation}
   \label{eq:eff_mass_trap}
   m^* (x)/m\approx 3\gamma (x)/2\pi^2=3 m g /2\pi^2\hbar^2 n(x).
 \end{equation}
Using the density profile (\ref{eq:dens_prof}) and introducing a complex function
\begin{equation}
  \label{eq:psi}
  n(x) \Phi(x,t)=\delta S_1 (x,t)+i \delta S_2 (x,t),
\end{equation}
one obtains from Eqs.~(\ref{eq:hydro_spin}):
\begin{equation}
  \label{eq:motion_psi}
  i\dot{\Phi}=\Omega\Phi =-\frac{\pi^2}{6}\frac{\omega}{\gamma_0 N}
  \frac{1}{\sqrt{1-X^2}}\partial_X\left(1-X^2\right) \partial_X \Phi,
\end{equation}
where $X=x/R$ is the dimensionless coordinate, and 
we assumed the stationary time dependence $\Phi (X,t)=
e^{-i\Omega t}\Phi(X)$.  Equation (\ref{eq:motion_psi}) shows that the
typical frequency scale of the isospin excitations is given by
$\omega/\gamma_0 N$, which is smaller than the scale $\omega$ of acoustic 
frequencies by a large factor $\gamma_0 N$. The exact solution to this 
equation was obtained numerically using the shooting method, and the 
spectrum shows only a small difference from the semi-classical result
 \begin{equation}
   \label{eq:semiclass}
  \Omega_j = \frac{A\omega}{\gamma_0 N}\left(j+\frac{1}{2}\right)^2,
   \qquad j=0,1,2,\ldots,
 \end{equation}
where the numerical factor is $A=\pi^5/48 \Gamma^4(3/4)\approx 2.83$. 
For $\omega \sim 100$~Hz, $\gamma_0 \sim 10$ and $N \sim 100$ as in the 
experiment \cite{Weiss}, the lowest eigenfrequencies $\Omega_j$ are two 
or three orders of magnitude smaller than acoustic frequencies and are 
$\sim 0.1$~Hz.

In conclusion, we have found extremely slow (iso)spin dynamics in the
strong coupling TG regime, originating from a very large effective mass
of spin waves. In an experiment with ultra-cold bosons, this should
show up as a spectacular isospin-density separation: an initial wave
packet splits into a fast acoustic wave traveling at the Fermi
velocity and an extremely slow spin wave \cite{Recati}. One can even 
think of ``freezing" the spin transport, which in experiments with 
two-component 1D Bose gases will correspond to freezing relative 
oscillations of the two components.

We thank F. Lalo\"e and W. Zwerger for useful discussions. This work was 
supported 
by the Minist\`ere de la Recherche (grant ACI Nanoscience 201), and by 
the Nederlandse Stichting voor Fundamenteel Onderzoek
der Materie (FOM). The studies were performed in part at Laboratoire 
Kastler Brossel, ENS Paris. LPTMS is UMR 8626 of
CNRS and Universit\'e Paris XI. LPS is UMR 8502 of CNRS and
Universit\'e Paris XI.


\begin{thebibliography}{99}
\bibitem{footnote} In order to excite an isospinon alone
we choose  density quantum numbers $I_j=j-N/2$ and the spin quantum 
number $J=1-N/2+l$,  where $l$ varies between $1$ and $N-2$ \cite{LGYE}.
\end{thebibliography}
\end{document}